\documentclass[runningheads]{llncs}

\usepackage[mobile]{eccv}

\usepackage{eccvabbrv}

\usepackage{graphicx}
\usepackage{booktabs}

\usepackage[accsupp]{axessibility}  

\usepackage{bm}
\usepackage{amsmath}
\usepackage{amssymb}
\usepackage{nicefrac}
\usepackage{dsfont}
\usepackage{colortbl}
\usepackage{comment}
\usepackage{caption}
\usepackage{wrapfig}
\usepackage{multirow,tabularx}

\usepackage{siunitx}
\usepackage{bm}
\usepackage{bbm}
\usepackage{sidecap}

\usepackage[pagebackref,breaklinks,colorlinks,citecolor=eccvblue]{hyperref}

\usepackage{orcidlink}

\begin{document}

\title{Generating Binary Species Range Maps} 


\author{
Filip Dorm\inst{1} \quad
Christian Lange\inst{1} \quad
Scott Loarie\inst{2} \quad
Oisin Mac Aodha\inst{1}
}

\authorrunning{F. Dorm \etal}

\institute{
$^1$University of Edinburgh \quad $^2$iNaturalist
}

\maketitle

\begin{abstract}
Accurately predicting the geographic ranges of species is crucial for assisting conservation efforts. Traditionally, range maps were manually created by experts. However, species distribution models (SDMs) and, more recently, deep learning-based variants offer a potential automated alternative. Deep learning-based SDMs generate a continuous probability representing the predicted presence of a species at a given location, which must be binarized by setting per-species thresholds to obtain binary range maps. However, selecting appropriate per-species thresholds to binarize these predictions is non-trivial as different species can require distinct thresholds. In this work, we evaluate different approaches for automatically identifying the best thresholds for binarizing range maps using presence-only data. This includes approaches that require the generation of additional pseudo-absence data, along with ones that only require presence data. We also propose an extension of an existing presence-only technique that is more robust to outliers. We perform a detailed evaluation of different thresholding techniques on the tasks of binary range estimation and large-scale fine-grained visual classification, and we demonstrate improved performance over existing pseudo-absence free approaches using our method. 
\end{abstract}

\section{Introduction}
\label{sec:intro}

Accurate range maps, depicting where species are present and absent, are indispensable tools for conservation. 
Statistics derived from their ranges are some of the key information used to estimate the threatened status of different species~\cite{standards2024guidelines}. 
However, traditional expert-derived range maps are costly and time consuming to create.
To alleviate this, advancements in statistics and machine learning have enabled the development of species distribution models (SDMs) that can estimate the probability of a species being present, among other quantities, at different locations~\cite{beery2021species}. 
Recently, deep learning-based multi-species SDMs, trained on crowd-sourced observation data, have been demonstrated to be capable of predicting the presence of thousands of different species simultaneously for an input location of interest~\cite{botella2018deep,mac2019presence,davis2023deep,lange2023active,teng2023bird,coleICML2023}. 
However, to obtain a binary range map from these continuous probabilities, they need to be binarized using a threshold. 
Binary range maps are used in visualization in addition to being necessary for computing spatial metrics for determining the threatened status of a species.

\begin{figure}[t]
\centering
\includegraphics[width=0.9\linewidth]{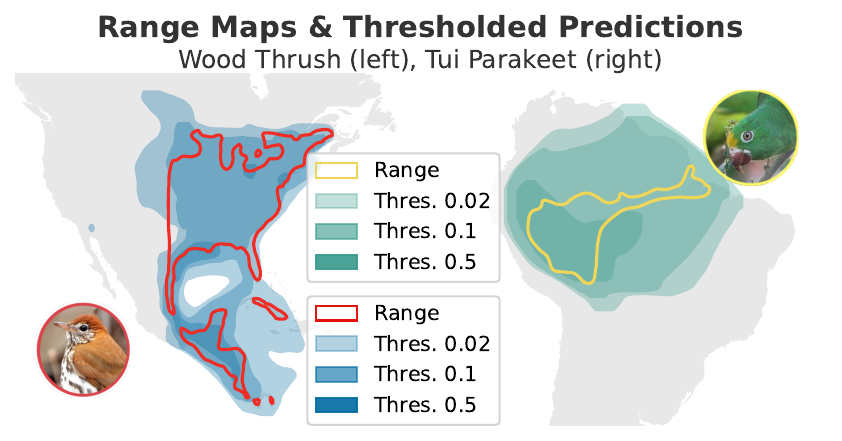}
\vspace{-12pt}
\caption{
{\bf Binary range maps for two different species.} 
These ranges are generated by the SINR~\cite{coleICML2023} species distribution model (SDM) for \href{https://www.inaturalist.org/taxa/13270-Hylocichla-mustelina}{Wood Thrush} (left) and \href{https://www.inaturalist.org/taxa/19208-Brotogeris-sanctithomae}{Tui Parakeet} (right), where the expert range map is denoted via solid outline. 
Converting the continuous SDM outputs to binary range maps requires setting thresholds (\eg $0.02$, $0.1$, or $0.5$) which result in very different range maps depending on the values chosen. 
More importantly, here the same threshold value is not the best for both species.  
}
\vspace{-15pt}
\label{fig:wood_range}
\end{figure}

Finding the most effective threshold is a challenging problem, as the one that leads to the most accurate range maps will not necessarily be the same across different species (see~\cref{fig:wood_range}). 
This is because the outputs of multi-species deep SDMs can be highly miscalibrated, which is likely the result of a number of factors such as imbalances in the species present in the training data, spatial biases, and issues stemming from species co-occurrences. 
The threshold selection task has been previously explored in the ecology literature under the setting where calibration data, which denotes a subset of locations where the species of interest is present or absent, is available~\cite{liu2005selecting}. 
However, the more challenging setting, and the one explored in this work, is where we do not have any information related to confirmed absences~\cite{liu2013selecting}. 
This presence-only data paradigm is most relevant to the opportunistic data available from large citizen science platforms such as iNaturalist~\cite{inat}, which contains over 200 million species observations to date.

A number of different thresholding approaches have been proposed for the case when only species presence observations are available~\cite{liu2013selecting}. 
However, many of these approaches require the generation of `pseudo-absences', \ie hallucinated data that simulates locations where a species is said to not be found. 
As a result, the precise design choices made when generating these pseudo-absences can have a large impact on the accuracy of the resulting binary range maps. 
In this work, we compare these approaches to alternative thresholding techniques that do not require absence data of any form. 
Furthermore, we outline a simple extension of these absence-free techniques that obtain superior performance. 
Our resulting approach is conceptually simple and more efficient than the more computationally expensive pseudo-absence-based alternatives.

We make the following contributions: 
(i) We perform a detailed evaluation of multiple different thresholding techniques for generating binary species range maps. Our experiments are conducted using recent multi-species SDMs over a large set of expert-derived evaluation range data.  
(ii) We show that binary range maps can also be used as priors to improve large-scale fine-grained image classification and that they are superior to existing continuous priors. 
(iii) Finally, we introduce an extension of an existing thresholding approach that does not require any pseudo-absence data yet still outperforms existing presence-only methods on both range estimation and image classification tasks. 
Code to reproduce our experiments is available here: \href{https://github.com/filipgdorm/binary_range_maps}{https://github.com/filipgdorm/binary\_range\_maps}

\vspace{-5pt}
\section{Related Work}
\vspace{-5pt}
\label{sec:rel_work} 

\noindent{\bf Species Distribution Models (SDMs).} There is a large body of work on developing models to estimate the spatial and temporal distributions of species from raw observation data~\cite{elith2009species,beery2021species}. 
These models are typically trained using one of two different types of data: presence-only (PO) or presence-absence data.
PO data consists of a set of observations (\ie locations in space and time where a species has been observed), without any confirmed absences.
PO data is significantly easier to collect compared to presence-absence data, requiring only the recording of locations where a species has been observed, without the need for extensive, resource-intensive surveys to confirm absences.
As a result, PO data is abundant and widely available through online citizen science initiatives such as iNaturalist~\cite{inat}. 

However, a challenge in training SDMs with PO data is the need for negative signal (\ie species' absences) during training. 
Without having absences in the training data, SDMs can naively learn to simply predict presence, for every species, at every location. 
One common approach to overcome this issue is to create \emph{pseudo}-absences (PAs), \ie artificially generated data points representing species' absences~\cite{barbet2012selecting}. 
Various approaches for generating PAs have been proposed, with popular techniques including selecting locations uniformly at random (\ie `random background points') or locations where other species have been observed (\ie `target background points')~\cite{ponder2001evaluation,phillips2009sample,botella2020bias}.
The effectiveness of different methods for generating PAs for training multi-species deep SDMs has recently been explored in~\cite{zbinden2024selection}.

Recently, \cite{coleICML2023} introduced a deep learning-based approach for modeling species' distributions, termed Spatial Implicit Neural Representations (SINRs). 
SINRs are parameterized as fully connected neural networks and can be trained on noisy crowd-sourced PO data. 
The authors demonstrated the ability to jointly estimate the ranges of thousands of species within a single model using PO training data.
They introduced new benchmarks for evaluating the accuracy of predicted range maps by comparing model outputs to expert-derived ranges. 
They also quantified the improvement in fine-grained visual classification for different species when using SDMs as priors to modify image classifier predictions based on the location where the image was captured~\cite{mac2019presence}.
However, one of the limitations of \cite{coleICML2023}'s range estimation evaluation is that it uses a continuous, \ie area under the curve, metric. 
As a result, the task of \emph{binary} range estimation is not directly evaluated as the evaluation protocol is agnostic to the choice of threshold used to produce a binary range map.
We use the models and data from~\cite{coleICML2023} for our evaluation and demonstrate that the threshold required can vary significantly across species and how to select the best one is not obvious.  

\noindent{\bf Binarizing Range Maps.}  
The output of SDMs such as~\cite{coleICML2023} are a set of continuous predictions representing species' presences at the evaluated locations. 
Binarizing these predictions to create binary range maps requires setting thresholds, which can result in very different maps depending on the thresholds selected. 
Furthermore, as seen in \cref{fig:wood_range}, the choice of best threshold can be species dependent.  
Without true absence data for calibration~\cite{fielding1997review,liu2005selecting}, the problem of converting continuous outputs to binary ranges, given only PO data, is challenging. 

The problem of selecting appropriate thresholds for SDMs has been previously explored in the ecology literature~\cite{liu2005selecting,liu2013selecting,liu2016selection}. 
However, these existing works tend to focus on an older generation of single species machine learning methods such as Maxent~\cite{phillips2006maximum}, Boosted Regression Trees~\cite{elith2008working}, and Random Forests~\cite{breiman2001random}. 
Additionally, these experiments are often limited to a small number of species, sometimes using only simulated data~\cite{liu2013selecting}, and constrained to smaller geographic regions. 
Previous work has shown that fixed constant thresholds (\eg 0.5) do not work well when the training data is imbalanced~\cite{jimenez2007threshold, freeman2008comparison}. 
In the context of PO data, the impact of different thresholding techniques has been evaluated in~\cite{liu2013selecting,liu2016selection}. 
Existing techniques can be broadly categorized based on the data they require, \eg no data, presences, or presences and PAs. 
When no data is available, approaches such as choosing a fixed threshold or simply setting the threshold to be the average of the predictions can be used. 
Alternatively, if presence data is available, one approach is to choose the lowest/minimum presence threshold based on the SDM predictions associated with the presence observations in the training data~\cite{phillips2006maximum,pearson2007predicting}. 
In essence, this approach attempts to maximize recall by minimizing the number of false negatives. 
Another approach is to generate continuous predictions for all test locations (\ie not just training presences) and choose the threshold that results in a fixed percentage of the lowest predictions being set as absences~\cite{pearson2007predicting}. 
However, the selected percentage is heavily influenced by the ratio between presence and absence data. 
If one generates PAs, then it is possible to select the threshold that directly maximizes the downstream evaluation metric of interest (\eg F1 score)~\cite{liu2013selecting} or to achieve a desired target value (\eg 95\% sensitivity)~\cite{pearson2004modelling}. 
However, these approaches are sensitive to mismatches between the distribution of true absences and generated PAs.  

We build on the above existing work by extending the evaluation of thresholding techniques beyond simulation data, and instead use a recent dataset of globally distributed expert-derived range maps of 3,000 different species from~\cite{coleICML2023}. 
Furthermore, our evaluation is based on recent multi-species neural-network-based SDMs that are jointly trained on PO data from thousands of species. 

\noindent{\bf SDMs as Geo Priors for Vision.} In addition to directly generating range maps, SDMs have also been shown to be helpful for assisting image classification. 
Several works have demonstrated that combining the probabilistic outputs of image classifiers with an estimate of the spatial distribution, \ie as a `geo prior', of the categories of interest can improve classification accuracy.
Additional image metadata, which encodes the geographic location where each image was captured, can be used to estimate the spatial distributions of different classes.
Once trained, the geo prior downweights the classifier predictions for categories that are unlikely to be present, according to the prior, at the location where the test image was taken. 
Different approaches have been explored in the literature, ranging from methods that spatially bin observations~\cite{berg2014birdsnap}, ones that jointly train classifiers with additional location inputs~\cite{tang2015improving,chu2019geo}, or methods that combine classifier predictions with separate spatial distribution models~\cite{mac2019presence}. 
In our later experiments we show that binary range maps provide a small, but non-trivial, improvement in classification accuracy compared to existing approaches.  

\vspace{-5pt}
\section{Method}
\label{sec:method}
\vspace{-5pt}
Our goal is to generate binary species range maps. 
We begin with an overview of how SDMs can be trained to generate continuous range maps, and then introduce different approaches for converting these continuous predictions to binary ones. 

\subsection{Predicting Species' Presence}
Recently~\cite{coleICML2023} introduced a neural network-based approach, SINR, for predicting continuous range maps.  
Their model is trained simultaneously on thousands of different species from crowd-sourced presence-only data.  
At test time, the input to the model is a longitude-latitude tuple denoting a location of interest, and the output is a multi-label classification indicating the probability that each of several thousand different species is present there.  
Below we provide an explanation of how their model is parameterized and subsequently trained.

SINR denotes an input location by its longitude and latitude as $\mathbf{x}_i = [lon, lat]$.  
As in~\cite{mac2019presence}, these inputs are wrapped using a sinusoidal encoding to limit the impact of boundary effects.  
Other more complex input transformations are possible, but this type of wrapped input is still competitive on related spatial representation learning tasks~\cite{wu2024torchspatial}. 
While it is possible to use other input features such as ones representing local environmental conditions, such as temperature and rainfall, in this work we mainly focus on location only models. 
The true absence (0) or presence (1) of $S$ different species at $\mathbf{x}_i$ is denoted by the multi-label vector $\mathbf{y}_i \in \{0, 1\}^S$. 
In practice, in the presence-only setting we do not have any training data that indicates if a species is absent. 
Instead, similar to~\cite{cole2021multi}, the observed training data is represented as $\mathbf{z}_i \in \{1, \emptyset\}^S$, where $z_{ij} = 1$ if a species is present and $z_{ij} = \emptyset$ if the presence or absence of the species is unknown. 
SINR models  estimate $\mathbf{y}_i$ at any location $\mathbf{x}_i$ within a spatial domain $\mathcal{X}$, given observational training data $\{(\mathbf{x}_i, \mathbf{z}_i)\}_{i=1}^N$. 

The range estimation model we wish to learn is denoted as $\hat{\mathbf{y}}_i = h_{\phi}(f_{\theta}(\mathbf{x}_i))$, where $f_{\theta}: \mathcal{X} \rightarrow \mathbb{R}^k$ is a location encoder parameterized by $\theta$, and $h_{\phi}: \mathbb{R}^k \rightarrow [0, 1]^S$ is a multi-label classifier parameterized by $\phi$. 
The location encoder $f_{\theta}$ is a fully connected neural network with residual connections and the species classifiers in $h_{\phi}$ are simple per-species linear projections followed by sigmoid non-linearities. 
The prediction $\hat{\mathbf{y}}_i \in [0, 1]^S$ represents a prediction of the probability of each species being present at location $\mathbf{x}_i$. 
The parameters of both components of the model can be estimated using stochastic gradient descent: 
\begin{equation}
\theta^*, \phi^* = \underset{\theta, \phi}{\text{argmin}} \frac{1}{N} \sum_{i=1}^{N} \mathcal{L}(\hat{\mathbf{y}}_i, \mathbf{z}_i),
\end{equation}
where $\mathcal{L}$ is a suitable loss function and $\hat{\mathbf{y}}_i = h_{\phi}(f_{\theta}(\mathbf{x}_i))$ is the model's prediction.  
The output of this model is thus a probability distribution indicating species' presence, or absence, over a location of interest.

\noindent{\bf Training Loss.} 
\cite{coleICML2023} evaluate different loss functions $\mathcal{L}$ for training  SINR models. 
At their core, these losses amount to different ways of addressing the lack of true absence data. 
To address this, different mechanisms for generating pseudo-absences (PAs) are explored. 
For example, taking inspiration from the task of single positive multi-label learning~\cite{cole2021multi}, the heuristic that all unobserved labels are negative is used. 
This assumption also holds for species range estimation because, given an arbitrary location, most species will not actually be present there. 
Here we focus on their best performing approach, `Full Assume Negative'  ($\mathcal{L}_{\text{AN-full}}$), which is inspired by losses introduced in~\cite{mac2019presence}. 
We describe the other losses in Appendix~\ref{sec:SINR_losses}.  

For the $\mathcal{L}_{\text{AN-full}}$ loss function, PAs are selected in two different ways. 
First, each species presence is paired with a PA at a location $\mathbf{r}$ that is sampled uniformly from the surface of the earth, \ie $\mathbf{r} \sim \text{Uniform}(\mathcal{X})$. 
Second, each species observation is also paired with a PA at the same location for a \emph{different} species, which generates PAs that align with the distribution of the data.  
The final loss is thus represented as:
\begin{equation}
\mathcal{L}_{\text{AN-full}}(\hat{\mathbf{y}_i}, \mathbf{z}_i) = -\frac{1}{S} \sum_{j=1}^{S} \left[ \mathbbm{1}_{[z_{ij}=1]} \lambda \log(\hat{y}_{ij}) + \mathbbm{1}_{[z_{ij} \neq 1]} \log(1 - \hat{y}_{ij}) + \log(1 - \hat{y}'_{j}) \right],
\end{equation}
where $\hat{y}_{ij}$ is the predicted probability of species $j$ being present, $z_{ij}$ indicates the presence or absence of species $j$, $\lambda > 0$ is a hyperparameter used to prevent PAs from dominating the loss, and $\hat{\mathbf{y}}' = h_{\phi}(f_{\theta}(\mathbf{r}))$ are the predicted probabilities of the species of interest being present at a randomly chosen location $\mathbf{r}$. 
This loss is efficient in that it provides a training signal for all entries in $\hat{\mathbf{y}}$ for each training example, where a positive signal is provided for the observed present species and a negative signal from PAs is provided for all other species.

%
%
%
\subsection{Generating Binary Range Maps via Thresholding} 
Thresholding techniques aim to estimate species-specific thresholds that lead to the most accurate binary range maps. 
Thresholds are applied after the SDM has been trained and after the predictions for each location have been generated. 
Once a threshold is identified, all predictions above or equal to it are marked as presences, and all below as absences. 
Next we outline different threshold generation techniques.  
\vspace{3pt}

\noindent{\bf Single Fixed Threshold.} This uses a single constant threshold (\eg 0.5) across all species~\cite{manel1999comparing, manel2001evaluating, li1997regression}. If the model is well calibrated, this should be sufficient.  

\noindent{\bf Single Best Threshold.} This method also uses a single threshold, but unlike the previous approach here we use the presence and absence data from the expert ranges in the evaluation data to select the threshold that leads to the highest mean F1 score. 
This threshold is selected by generating 20 linearly spaced candidate thresholds in the range 0-1 and choosing the best one. 

\noindent{\bf Random Sampling.}
\label{random_method} 
When true absences are not available, one common choice is to generate pseudo-absences (PAs) \cite{coleICML2023, zbinden2024selection}. 
This method takes presence observations from the \emph{training} data and generates randomly distributed PAs for each species where it has not been observed and selects per-species thresholds that maximize the mean F1 score on this data. This is achieved by binning the observations into H3~\cite{H3Web} cells of resolution four. The cells chosen as presences are those where the species has been observed, the absences are randomly sampled from the rest of the cells. In our later evaluation, we explore the impact of varying the ratio between the true presence and generated PA data. 

\noindent{\bf Target Sampling.}  
\label{method_tgt} 
To match the distribution of the training data, this approach generates PAs in locations where other species have been observed, but the species of interest is absent \cite{coleICML2023, zbinden2024selection}.
We use H3~\cite{H3Web} cells of resolution four to bin the presence data, where cells with at least one observation of a species are counted as presences.  
All cells where there are at least $N_{t}$ observations across all species, but the species of interest is not present, are counted as PAs. 
The value of $N_{t}$ is calculated by first identifying all the cells where the species of interest has been observed at least once. These cells are ranked based on the ratio of the total number of observations in each cell to the number of observations specifically for the species of interest within that cell. $N_{t}$ represents the 95th percentile value of this ratio, indicating the minimum number of total observations in a cell needed to be reasonably confident that the species of interest is not present in that cell.
Given the resulting presence and PA data, we can select per-species thresholds that maximize the F1 score.

\noindent{\bf Threshold Classifier.}
For this method we introduce a novel supervised machine learning approach to predict the best threshold per-species.  
For the input features for each species we use the final learned species embedding vector from~\cite{coleICML2023}, where the assumption is that species with similar embeddings will require similar thresholds. 
For the output of the learning algorithm, we bin the thresholds into 20 possible values, evenly distributed between 0-1, and treat it as a classification problem. 
75\% of the 2,418 IUCN species are used to train the classifier and the remaining 25\% are used for evaluation. 
As a result, the final scores for this approach are not directly comparable to the other techniques. 
We explore two different variants of this approach: a Random Forest (RF) classifier which uses the default values from scikit-learn~\cite{pedregosa2011scikit} or a Multilayer Perceptron (MLP) with two hidden layers, containing 100 neurons each. 

\noindent{\bf Mean Predicted Threshold.}
Following the approach outlined in~\cite{liu2013selecting}, we select the threshold based on the average of the SDM predictions. Specifically, for each species, we generate predictions across all locations. Subsequently, we determine the threshold by calculating the mean predicted value from a randomly selected subset of these locations.

\noindent{\bf Lowest Presence Threshold (LPT).}
This approach extracts the continuous predictions for a species of interest associated with each presence observation from the training data and selects the threshold as the lowest value from this set~\cite{pearson2007predicting}. 
This then ensures that the resulting binary range is guaranteed to encompass each of the presence observations.

\noindent{\bf Lowest Presence Threshold - Robust (LPT-R).}
\label{masking_method}
This method builds on the previous \texttt{LPT} approach~\cite{pearson2007predicting}, but we extend it to make it more robust to outliers. 
The motivation is that \texttt{LPT} is highly susceptible to outliers as inaccurate observations or `vagrants' (individuals lost outside their ranges~\cite{lees2022vagrancy}) can massively bias the threshold selection. 
So instead of choosing a threshold that includes all training presences, we instead select one that encompasses the vast majority of them. 
Specifically, we perform the same steps as \texttt{LPT} but set the threshold by sorting all the predicted continuous scores and choosing it to be equal to the 5th percentile. 
\cref{fig:masking_fig} in the appendix illustrates a visual example of this method.

\vspace{-5pt}
\section{Implementation Details}
\vspace{-5pt}

\subsection{Datasets}
\label{sec:eval_datasets}

\noindent{\bf Presence-Only Training Data.}
Our training data is publicly available crowd-sourced occurrence data from iNaturalist~\cite{inat}. 
The original raw data contains annotated images of species with geographic locations and timestamps. 
We use the same subset of 35.5 million total observations containing 47,375 species from~\cite{coleICML2023}. Here, observations are of `research grade' status, meaning that there is a higher confidence that the classifications are correct. 
While iNaturalist offers numerous advantages, it also has limitations. 
For instance, the data contained within is biased towards the Global North. 

\noindent{\bf Presence-Absence Evaluation Data.} 
For evaluation, we require data that contains both confirmed presence \emph{and} absence observations distributed across the globe.  
The S\&T and IUCN datasets from~\cite{coleICML2023} satisfy these requirements. 
The \textit{eBird Status and Trends} (S\&T) dataset covers 535 bird species biased to North America~\cite{Fink2020}. This dataset contains estimated relative abundance maps which have been converted into binary range maps.
The second dataset consists of ranges of 2,418 species from the \textit{International Union for Conservation of Nature} (IUCN)~\cite{iucnredlist2022}. In contrast to the S\&T dataset, this dataset is more taxonomically diverse and more globally distributed. 
The two evaluation datasets are represented as H3~\cite{H3Web} cells at resolution five, which corresponds to 2,016,842 evaluation locations distributed across the entire globe. 

\noindent{\bf Image Classification Evaluation Data.} 
In addition to range estimation, we also evaluate different models as geographic priors on the task of assisting image classification predictions as in~\cite{berg2014birdsnap,mac2019presence,coleICML2023}.  
We again use the same dataset from~\cite{coleICML2023}, which contains 282,974 images from iNaturalist, which represents a subset of 39,444 species from our training set. 
The evaluation metric for this task is the top-1 image classification accuracy resulting from combining, in our case binary, range predictions with the probabilistic classification outputs of an Xception classifier~\cite{chollet2017xception} trained on images from iNaturalist.

\subsection{Evaluation Metrics}
\label{sec:eval_metrics} 
Continuous range predictions can be evaluated using area under the curve-based metrics such as average precision. 
These scores can then be averaged across different species to come up with a single final score as in~\cite{coleICML2023}. 
However, our goal is to evaluate binary range maps, not continuous predictions.  
Thus after thresholding a continuous probability of presence output from a model, we consider all locations above the threshold as representing predicted presences and all below as predicted absences. 
Evaluation is performed using the F1 score, which combines precision and recall into a single value and is particularly useful in binary classification problems where there is an imbalance between the classes, which is the case here as the presences are typically outnumbered by absences. 
The F1 score is computed per-species after binarization:
\begin{equation}
    F_1 = 2 \cdot \frac{\text{precision} \cdot \text{recall}}{\text{precision} + \text{recall}} = \frac{\text{TP}}{\text{TP} + \frac{1}{2}(\text{FP} + \text{FN})},
\end{equation}
where TP denotes True Positives, FP False Positives, and FN False Negatives. 
The final reported mean F1 score is the average per-species F1 score across the species in each of the respective evaluation sets. 
Unless stated otherwise, the held-out evaluation datasets are completely detached from any threshold identification methods to ensure that there is no data leakage in the process.

\vspace{-5pt}
\section{Experiments}
\vspace{-5pt}
\label{sec:experiments}
In this section we evaluate different techniques for generating binary range maps. 
Unless specified otherwise, experiments are carried out with the $\mathcal{L}_{\text{AN-full}}$ SDM from~\cite{coleICML2023} capped at 1000 training samples per-species without environmental input features. 
The model's predictions are binarized using different thresholding techniques and then evaluated by computing the mean F1 score against the held-out test set. 
This score is presented together with information about how far the performance is from the upper bound. 
We also provide additional quantitative results, including different input features, and analysis in \cref{sec:additional_results}.

\begin{table*}[t]
    \centering
    \caption{{\bf Binary range estimation performance of different thresholding techniques.} 
    Here we report the average mean F1 score for five different random initializations of the $\mathcal{L}_{\text{AN-full}}$ SDM on the IUCN evaluation set, where the upper bound is 67.2\%. ($\dagger$) denotes threshold classifier scores which are computed on a 25\% subset of the full evaluation set, as they are trained on the rest. 
    (\checkmark) indicates whether a thresholding technique uses true absences (TA), pseudo-absences (PA), or one single overall threshold (ST). 
    Bold entries indicate best methods, and underline are second best. 
    }
    \vspace{-7pt}
    \resizebox{0.95\textwidth}{!}{
    \begin{tabular}{l|ccc|cc}
    \toprule
    Thresholding Method & ST& PA & TA & $\uparrow$ Mean F1 & $\downarrow$ Upper Bound~$\Delta$\\
    \midrule 
     Threshold Classifier - RF $^\dagger$      & & & \checkmark &55.4   &$-11.8$ \\
     Threshold Classifier - MLP $^\dagger$     & & & \checkmark &56.3   &$-10.9$\\ \hline
     Single Fixed Threshold - 0.5        & \checkmark & & &40.2   &$-27.0$\\
     Single Best Threshold - 0.1     & \checkmark & & \checkmark &55.2   &$-12.0$ \\
     Random Sampling - \#Absences=\#Presences            & &\checkmark & &57.6   &$-9.6$ \\
     Random Sampling - \#Absences=5\#Presences           & &\checkmark & &58.1   &$-9.1$ \\
     Random Sampling - \#Absences=10\#Presences          & &\checkmark & &56.9   &$-10.3$ \\
     Random Sampling - 100 Absences                      & &\checkmark & &57.3   &$-9.9$ \\
     Random Sampling - 1000 Absences                     & &\checkmark & &57.6   &$-9.6$ \\
     Random Sampling - 10000 Absences                    & &\checkmark & &43.9   &$-23.3$ \\
     Target Sampling            & &\checkmark & &\underline{59.7}   &\underline{$-7.5$} \\
     Mean Predicted Threshold       & & & &37.7   &$-29.5$\\
     Lowest Presence Threshold (LPT) & & & & 54.3  &$-12.9$\\
     Lowest Presence Threshold - Robust (LPT-R)   & & & &\textbf{60.8}   &$-$\textbf{6.4} \\
    \bottomrule
    \end{tabular}
    }
    \label{tab:more_results}
    \vspace{-10pt}
\end{table*}

\vspace{-5pt}
\subsection{Binarizing Range Maps} 
\noindent{{\bf Evaluating Different Thresholding Techniques.}} \cref{tab:more_results} presents the mean F1 score for a wide set of binarization techniques when evaluated on the IUCN dataset from~\cite{coleICML2023}. 
The robust Lowest Presence Threshold (\texttt{LPT-R}) method is the highest-performing technique, closely followed by the \texttt{Target Sampling} approach. 
Both the \texttt{LPT} and \texttt{LPT-R} methods have the advantage that they do not require any pseudo-absences to be defined. 
Neither \texttt{Threshold Classifier} techniques work particularly well here when predicting thresholds, indicating that the relationship between SDM species' weights and final thresholds is not very effective for this type of model. 
Still, these methods, along with the vast majority of techniques, work better than using one \texttt{Single Fixed Threshold} of 0.5 for binarization. 
The main exception to this is the \texttt{Mean Predicted Threshold} technique which performs worse than this baseline. 
Unlike~\cite{liu2013selecting}, which highlights this approach as viable in the PO setting, our evaluation on a larger set of globally distributed species indicates that it is not very effective. 
Generally the SDM outputs are very small because most species are absent from most places. 
Thus, selecting the threshold by averaging predictions results in a small threshold.

\begin{figure}[th]
\centering
\includegraphics[width=0.97\textwidth]{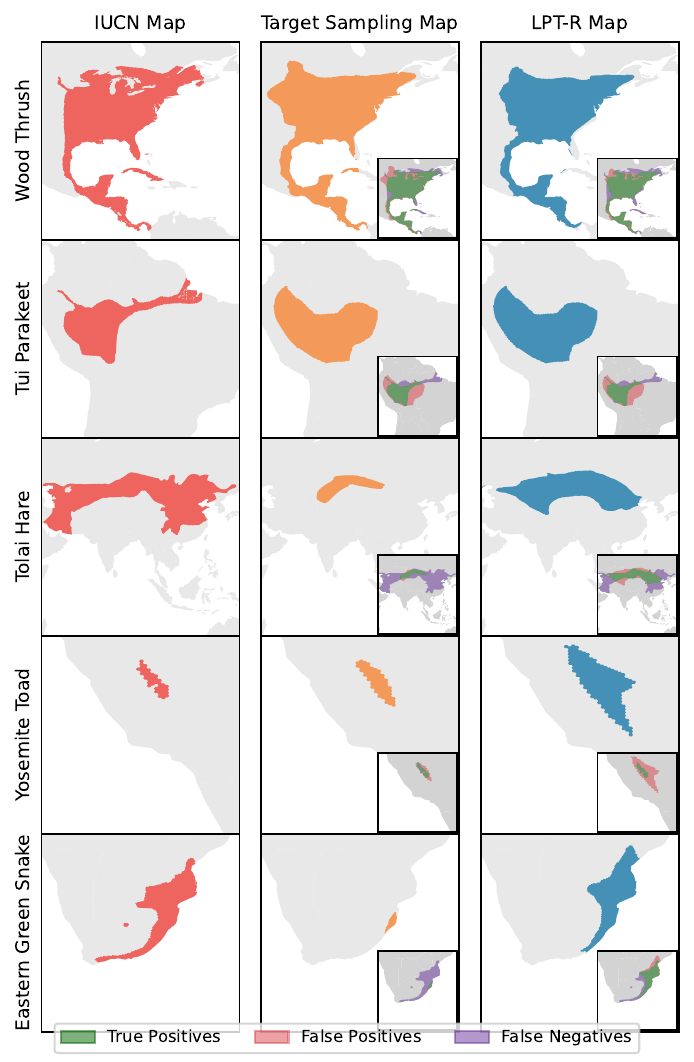}
\vspace{-10pt}
\caption{{\bf Qualitative examples of estimated binary ranges}. 
Each row depicts a different species, and the columns show the expert-derived ranges and the outputs from the \texttt{Target Sampling} and \texttt{LPT-R} approaches, respectively. Inset, we also display the different types of errors. We use an ocean mask for visualization purposes. 
}
\label{fig:qualitative_results}
\end{figure}

Qualitative examples of the resulting binary ranges for several species are depicted in \cref{fig:qualitative_results}.  
This figure displays a representative set of species and highlights how the two best-performing methods compare to expert-derived range maps. 
In many cases, such as the \emph{Wood Thrush} and \emph{Tui Parakeet}, the methods perform similarly. 
The former case is an example of a North American bird, a species that has abundant PO training data. 
The latter is a South American species with much less training data, but where the two methods yield almost identical threshold and range maps. 
The \emph{Tolai Hare} and \emph{Eastern Green Snake} are both examples where the \texttt{LPT-R} approach generates much more accurate ranges. 
Both of these species are in regions with less training data and have large ranges, as seen in expert ranges on the left. 
In these examples, the \texttt{Target Sampling} greatly underestimates the ranges. 
This seems to often be the case for species with large ranges and few training samples. 
However, for species like the \emph{Yosemite Toad} with few training samples and a small range, the \texttt{Target Sampling} approach's more conservative range estimation appears to work better.

\noindent{{\bf Evaluating Different SDMs.}} Next, in \cref{tab:big_results_table} we evaluate the best-performing techniques more using different input SDMs and additional datasets to quantify their robustness.  
Here, the different SDMs correspond to training SINR models with one of three different loss functions. 
We include results on the smaller S\&T dataset from~\cite{coleICML2023}, which is biased toward birds from North America with larger ranges. 
As can be seen in the results, for the more challenging IUCN dataset, the \texttt{LPT-R} approach outperforms the other methods for all three input SDMs. 
The \texttt{Target Sampling} approach is the second best-performing method for $\mathcal{L}_{\text{AN-full}}$ and $\mathcal{L}_{\text{AN-SSDL}}$, while being the second worst method for $\mathcal{L}_{\text{AN-SLDS}}$. 
The performance gap is at most a few percentage points, but considering that the \texttt{LPT-R} approach foregoes several computation steps associated with PA generation used by the other methods, this highlights its utility. 
As \texttt{LPT-R} performs well on the IUCN dataset, it perhaps implies that it may work better for a wider set of species, even those that are less prevalent on iNaturalist. 
This is important, given that these types of species are often the ones that require monitoring.

For the S\&T evaluation dataset, the \texttt{Target Sampling} approach generally performs best. However, for the $\mathcal{L}_{\text{AN-SLDS}}$ SDM, the performance of the different top techniques is almost identical. 
As noted earlier, the S\&T dataset has a North American bias, which is also where most of the iNaturalist training data is from. 
Similar to the findings in~\cite{coleICML2023}, the  $\mathcal{L}_{\text{AN-SLDS}}$ SDM performs well on the S\&T dataset, which is likely due to the fact that this loss function uses a version of target background sampling when generating its pseudo-absences which will also result in a bias towards North America in the context of this dataset.

\begin{table*}[t]
    \centering
    \caption{{\bf Binary range estimation across different input SDMs.} 
    Here we report the mean F1 score for three different input SDMs, where each is trained with a different loss function. 
    The reported scores are the result of five repeated runs, with different initializations, for each SDM. 
    We also report results for an ensemble of the five different $\mathcal{L}_{\text{AN-full}}$ SDMs, where we average the model predictions before thresholding. 
    }
    \vspace{-7pt}
    \resizebox{0.95\textwidth}{!}{
    \begin{tabular}{l|l|cc|cc}
    \toprule
    & & \multicolumn{2}{c|}{\textbf{S\&T}} & \multicolumn{2}{c}{\textbf{IUCN}} \\
    \toprule
    Model    & Method & $\uparrow$ Mean F1 & $\downarrow$ Upper Bound~$\Delta$  & $\uparrow$ Mean F1 & $\downarrow$ Upper Bound~$\Delta$\\
    \midrule
    \multirow{4}{*}{$\mathcal{L}_{\text{AN-full}}$}  
                                    &Random Sampling     &67.4 &$-8.7$       &57.6   &$-9.6$ \\
                                    &Target Sampling     &\textbf{70.5} &$-$\textbf{5.6}       &\underline{59.7}   &\underline{$-7.5$} \\
                                    &LPT &60.0&$-16.1$&54.3&$-12.9$\\
                                    &LPT-R   &\underline{69.1}&\underline{$-7.0$}  &\textbf{60.8}   &$-$\textbf{6.4} \\
                                    
    \hline
    \multirow{4}{*}{$\mathcal{L}_{\text{AN-SSDL}}$}  
                                    &Random Sampling     &57.8 &$-11.9$       &50.5   &$-9.6$ \\
                                    &Target Sampling     &\textbf{64.1}&$-$\textbf{5.6}       &\underline{53.1}   &\underline{$-7.0$} \\
                                    &LPT &58.2&$-11.5$&49.9&$-10.2$\\
                                    &LPT-R   &\underline{61.3} &\underline{$-8.4$}  &\textbf{53.5}   & $-$\textbf{6.6} \\
    \hline
    \multirow{4}{*}{$\mathcal{L}_{\text{AN-SLDS}}$}  
                                     &Random Sampling     &\textbf{71.1}&$-$\textbf{5.1}      &\underline{38.9}   &\underline{$-8.5$} \\
                                     &Target Sampling     &71.0&$-5.2$      &37.4   &$-10.0$\\
                                     &LPT &59.5&$-16.7$&30.0&$-17.4$\\
                                     &LPT-R  &\textbf{71.1}& $-$\textbf{5.1} &\textbf{39.4}   & $-$\textbf{8.0} \\
                                   
    \hline
    \multirow{4}{*}{\shortstack{Ensemble\\$\mathcal{L}_{\text{AN-full}}$}}
                                            &Random Sampling &67.5 & $-9.4$ &58.3 &$-10.0$\\
                                            &Target Sampling &\textbf{71.0} & $-$\textbf{5.9} &\underline{60.5} &\underline{$-7.8$}\\
                                            &LPT &62.8&$-14.1$&56.2&$-12.1$\\
                                            &LPT-R           &\underline{70.0} & \underline{$-6.9$} &\textbf{61.7} &$-$\textbf{6.6}\\
    \bottomrule
    \end{tabular}
    }
    \label{tab:big_results_table}
    \vspace{-15pt}
\end{table*}

\noindent{{\bf Ensembling SDMs.}} Finally, in the bottom rows of \cref{tab:big_results_table} we study the impact of ensembling five different random initializations of the $\mathcal{L}_{\text{AN-full}}$ SDM. 
This ensemble was used for both threshold identification and, subsequently, evaluation on both S\&T and IUCN datasets. 
The ensemble results in more accurate range maps compared to a single model, as illustrated by the higher mean F1 score for both datasets.
We also observe a reduction in the difference between \texttt{LPT-R} and \texttt{Target Sampling} on S\&T as a result of ensembling.

\vspace{-7pt}
\subsection{Geo Priors for Image Classification}
\vspace{-5pt}
SDMs can also be used as `geo priors' to assist image classifiers by combining their outputs with a classifier's probabilistic predictions. 
Here, we conduct an evaluation similar to the one carried out in~\cite{coleICML2023}, with the main difference being that with binary range maps, locations outside a species range would be downweighted to zero, \ie the SDM would prevent the classifier from predicting that species. 
We use the image classification dataset from~\cite{coleICML2023}, which consists of 282,974 images of different species collected from iNaturalist, covering 39,444 species from the PO training set.  
Each image is accompanied by the latitude and longitude at which it was taken. 
In \cref{tab:vision_results}, the results of using different binary range maps are presented. Using a geo prior to weight predictions increases the Top-1 accuracy in all experiments compared to the vision-only prediction model (75.4\%), except when binarizing with a constant threshold of 0.5 for all species. 
As previously established, different species require different thresholds, and a threshold of 0.5 is generally too high. This results in most species being incorrectly marked as absent from nearly all locations, which degrades model performance. 
However, setting a lower threshold of 0.1 for binarizing all species significantly improves accuracy, and outperforms the vision-only model. 
This shows that the vision model benefits from a geo prior, even if the range maps are not very accurate when making predictions. 

Switching to more refined methods of setting species-specific thresholds improves the performance even further. 
Out of all methods, the \texttt{Target Sampling} and \texttt{LPT-R} approaches work best. 
However, \cref{tab:vision_results} shows that this performance is lower than when using the raw prediction scores without binarizing them, \ie the original continuous outputs of the SDM (81.6\%). 
We hypothesize that the continuous scores allow the model to express uncertainty about a species being at a given location rather than choosing between the two extremes of classifying as either present or absent. 
To address this, we add a small constant $+\delta$ to the predicted absences after thresholding when using them as priors, \eg 0.01. 
In all cases, adding this constant improves the performance of all binarization techniques as it prevents them from saying that a species is never present at a location. 
As a result, the binarized range maps generated through both \texttt{Target Sampling} and \texttt{LPT-R} outperform the continuous baseline.

\begin{SCtable}[50][t]
    \centering
    \resizebox{0.43\textwidth}{!}{
    \begin{tabular}{l|c|c}
     \toprule
     Thresholding Method & \multicolumn{2}{c}{Top-1 Acc.}  \\
     \midrule
     & & +$\delta$ \\
    \midrule
    No Prior (\ie vision-only) &75.4 & -\\
    \hline
    Continuous - No Threshold & 81.6 & 81.4 \\
    Single Fixed Threshold - 0.5 &52.7 & 75.7\\
    Single Fixed Threshold - 0.1 &77.6 & 81.2\\
    Single Fixed Threshold - 0.01 & 80.9 & 81.3\\
    Random Sampling  & 78.5 & 81.5\\ 
    Target Sampling  &80.1 & \underline{81.7}  \\
    LPT  & 80.4 & 80.8  \\
    LPT-R & 80.0  & \textbf{81.8}   \\    
    \bottomrule
    \end{tabular}
    }
\caption{ {\bf Image classification with geo priors using different thresholding techniques.} 
Here we use a $\mathcal{L}_{\text{AN-full}}$ SDM as a geo prior to assist image classification. 
The results are the average of five different random initializations of the SDM.
`+ $\delta$' indicates that a small constant is added after thresholding to ensure that it is not possible for a thresholding technique to predict that a species is never present. 
}
\vspace{-12pt}
\label{tab:vision_results}
\end{SCtable}

\vspace{-7pt}
\subsection{Limitations} 
\vspace{-7pt}
Despite the promise of our work, there are still some important limitations to note. 
First, the training data contains biases, both spatial biases at the national (\eg Europe and North America are disproportionately represented) and local scale (\eg it is biased towards locations that are easier for people to get to), in addition to some species being much more prevalent than others. 
Spatial biases in citizen science data have been well characterized in the literature~\cite{geldmann2016determines} and various approaches have been proposed to mitigate it~\cite{chen2019bias,johnston2020estimating}. 
We leave the investigation of such methods for future work. 
Another data related issue is that the expert-derived range maps in the test set can contain errors or simply be outdated~\cite{coleICML2023}. 
However, to the best of our knowledge the test set used represents the best source of large-scale global evaluation data available.  
The main results in this work use species distribution models that only use coordinate features as input. 
Results can change if additional environmental covariates are used as input, see \cref{sec:additional_results}.  
We also do not perform any regularization on the binary maps to make them more spatially contiguous. 
This could be investigated in future work, although we note that our predictions are typically already contiguous. 
Finally, caution should be exercised when using the resulting binary range maps in conservation or assessment applications as the results demonstrate that performance is still likely insufficient for many species.

\vspace{-7pt}
\section{Conclusion}
\vspace{-7pt}
We explored the problem of automatically converting continuous species distribution model outputs into binary range maps via thresholding. 
Through detailed evaluation, we compared the performance of multiple different thresholding techniques, using different underlying models, across a range of different species. 
We also proposed an extension of an existing method, which we call \texttt{LPT-R}, and demonstrate that leads to more accurate binary range maps in a number of cases. 
Our approach circumvents the need to create pseudo-absences, making it computationally cheaper while also yielding more accurate ranges.  
Additionally, we showed that binary range maps can also be used as geographic priors to improve fine-grained image classification accuracy. 
A possible future extension to our work would be to leverage temporal data to generate time-conditioned range maps, which could significantly enhance our understanding of distribution changes over time.  

\vspace{5pt}
\noindent{\bf Acknowledgments.} We wish to thank the iNaturalist community for sharing their observation data. Thanks also to Alison Johnston for helpful comments. FD and OMA were in part supported by a Royal Society Research Grant.

%
%
\clearpage
\bibliographystyle{splncs04}
\bibliography{main}

\clearpage
\appendix
\setcounter{table}{0}
\renewcommand{\thetable}{A\arabic{table}}
\setcounter{figure}{0}
\renewcommand{\thefigure}{A\arabic{figure}}
\noindent{\LARGE \bf Appendix}

\vspace{-5pt}
\section{Additional Results}
\vspace{-5pt}
\label{sec:additional_results}
In this section, we present additional results and analysis. 
Unless stated otherwise, to generate these results we used the outputs from a single $\mathcal{L}_{\text{AN-full}}$ SDM and binarized the output using the \texttt{LPT-R} approach. 

\begin{figure}[h!]
    \centering
        \includegraphics[width=0.7\linewidth]{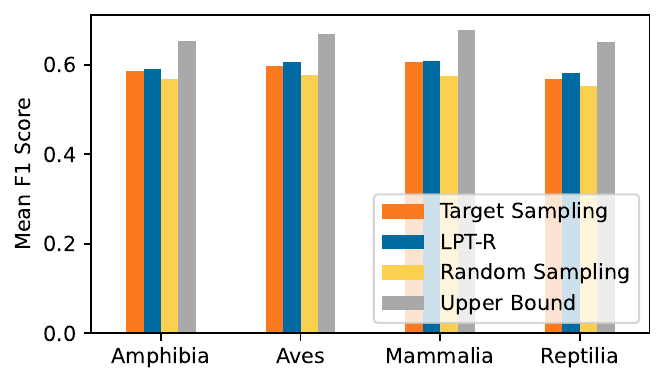}
    \vspace{-10pt}
    \caption{{\bf Results across different taxonomic groups.} Performance of the $\mathcal{L}_{\text{AN-full}}$ model on the IUCN task presented as the mean F1 score per taxonomic group.
        }
    \vspace{-10pt}
        \label{fig:iconic}
\end{figure}

\noindent{\bf How much does the performance vary for different taxonomic groups?} As seen in \cref{fig:iconic} for the IUCN dataset, \texttt{LPT-R} outperforms the other two approaches for the four different coarse taxonomic classes: amphibians, birds, mammals, and reptiles. 
This indicates that the results are stable across widely different taxonomic groups.

\begin{figure}[h!]
    \centering
        \includegraphics[width=0.7\linewidth]{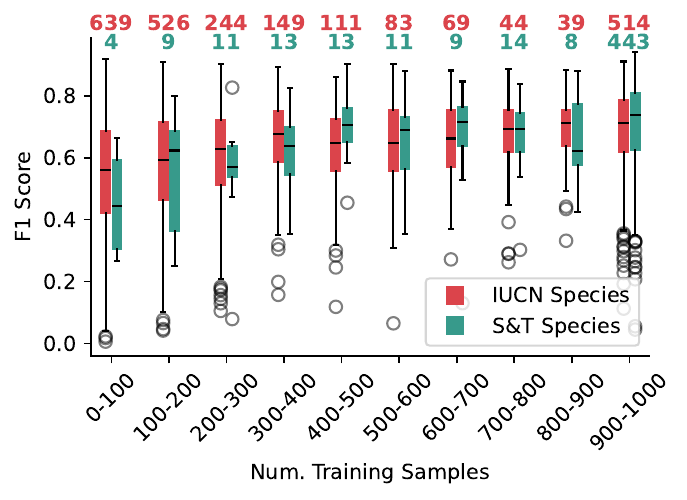}
    \vspace{-10pt}
    \caption{{\bf Performance against number of training examples.} Here we group species depending on how many training presence observations they have. 
    The number of species for each bin is written on top of each box plot. 
    The F1 score is calculated for the \texttt{LPT-R} method and the results are reported separately for the IUCN and S\&T datasets.     
    In general, performance improves with the number of training observations. 
    }
    \vspace{-10pt}
        \label{fig:app_fig_samples_v_score}
\end{figure}

\noindent{\bf How does the number of training samples influence the quality of binarized ranges?} 
In \cref{fig:app_fig_samples_v_score} we display the relationship between the number of training samples per species against the F1 score, \ie the measure of quality of the predicted binary range  maps. 
Results are reported separately for the IUCN and S\&T datasets. 
The overall trend is that the F1 score increases together with the number of training samples, \ie species with more training presence observations have better predicted ranges.

\begin{figure}[h!]
    \centering
        \includegraphics[width=0.7\linewidth]{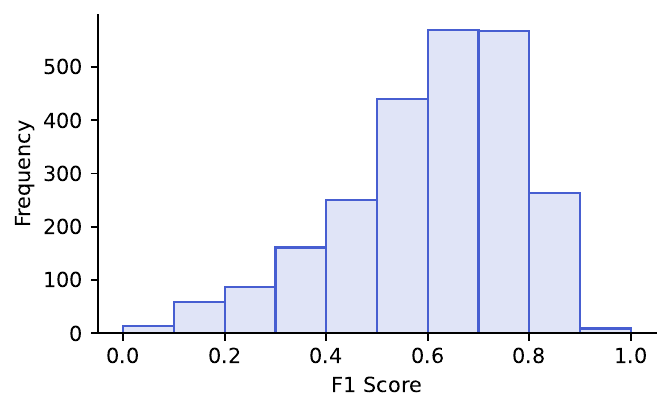}
    \vspace{-10pt}
    \caption{{\bf Per-species binned performance.} Histogram of scores on the IUCN dataset for $\mathcal{L}_{\text{AN-full}}$ binarized using \texttt{LPT-R}. 
    The x-axis represents binned F1 score, and the y-axis is the number of species in each bin. 
    In general, we observe that the distribution is skewed to the right.  
    }
    \label{fig:fscore_hist}
    \vspace{-10pt}
\end{figure}

\noindent{\bf What is the distribution of F1 scores across species?} 
In \cref{fig:fscore_hist} we display a histogram for the F1 scores for \texttt{LPT-R} on the IUCN dataset. 
We can see that most species obtain an F1 score of between 0.6 and 0.7 and that the distribution is skewed to the right. 
This indicates that thresholding results in plausible binary range maps for most species.  

\begin{figure}[h!]
    \centering
        \includegraphics[width=0.7\linewidth]{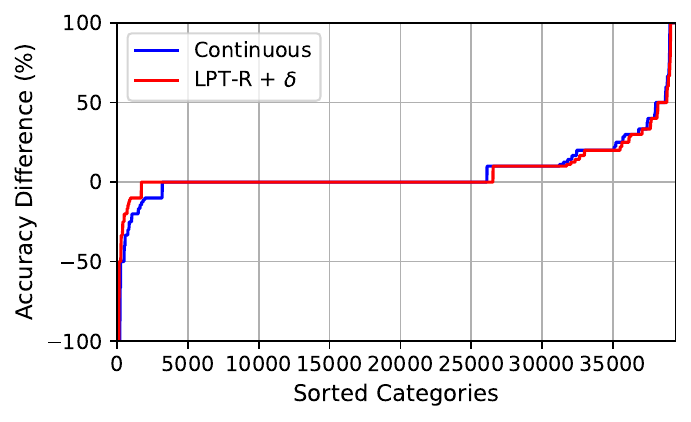}
    \vspace{-10pt}
    \caption{{\bf Per-species image classification performance improvement.} Here we sort species (\ie categories) in the fine-grained image classification task according to how much the classification accuracy improves after using an SDM as a geo prior. 
    Thus an accuracy difference of 0, indicates that the prior does not help for that particular species. 
    The sorted accuracies after applying the original continuous predictions as a geo prior are shown in blue. 
    Red shows the improvement resulting from using a binarized range map from our \texttt{LPT-R} with a small $\delta$ added. 
    } 
    \label{fig:geo_prior_gain}
    \vspace{-10pt}
\end{figure}

\noindent{\bf How many species obtain a boost in image classification performance as a result of using a binary range geo prior?}   
In \cref{fig:geo_prior_gain} we illustrate how using different geo priors influences the classification accuracy for computer vision models. 
We see that compared to the baseline of using continuous SDM predictions as a prior, the binarized range map results in fewer species with reduced performance as a result of using a prior (see left side of plot). 

\noindent{\bf How well do the thresholding techniques perform with models trained with additional environmental input features?}
In addition to the experiments in the main paper where we only evaluate SDMs that use coordinate features as input, here we evaluate SINR models that are trained using both environmental features and coordinates. 
SINR showed that these combined features yield the best continuous range estimation performance.
Our results for binarizing these outputs with different thresholding techniques can be found in \cref{tab:more_results_env}. 
The upper bound, \ie the theoretically best possible range maps, for the model with environmental features is substantially higher at 73\% (\vs 67.2\% for the coordinate-only case). 
However, the resulting binarized range maps are not much more accurate than those of the coordinate-only variant. 
For the environmental model, the \texttt{Target Sampling} method outperforms the \texttt{LPT-R} method. This is consistent with earlier experiments where \texttt{Target Sampling} was superior in the S\&T task, a set of species with more training data for which we expect the model to generate more accurate range maps. As the higher upper bound shows, there is potential for more accurate range maps, and this increased accuracy allows \texttt{Target Sampling} to outperform our \texttt{LPT-R}. The threshold classifier using an MLP also performs well, indicating a stronger correlation between model weights and optimal thresholds for certain model types. However, since the scores for this method are calculated using only a 25\% subset of species from the evaluation set, conclusions drawn from these results should be approached with caution.

\begin{table*}[t]
    \centering
    \caption{{\bf Binary range estimation performance of different thresholding techniques for models using additional environmental features.} 
    Similar to Tab.~\textcolor{red}{1}, here we report the average mean F1 score for five different random initializations of the $\mathcal{L}_{\text{AN-full}}$ SDM on the IUCN evaluation set. However, here the models are trained with coordinate \emph{and} environmental input features,  where the upper bound is 73.0\%. ($\dagger$) denotes threshold classifier scores which are computed on a 25\% subset of the full evaluation set, as they are trained on the rest, and thus cannot be directly compared. 
    (\checkmark) indicates whether a thresholding technique uses true absences (TA), pseudo-absences (PA), or one single overall threshold (ST). 
    Bold entries indicate best methods, and underline are second best. 
    }
    \vspace{-7pt}
    \resizebox{0.92\textwidth}{!}{
    \begin{tabular}{l|ccc|cc}
    \toprule
    Thresholding Method & ST& PA & TA & $\uparrow$ Mean F1 & $\downarrow$ Upper Bound~$\Delta$\\
    \midrule 
     Threshold Classifier - RF $^\dagger$      & & & \checkmark     &60.5 &$-12.5$\\
     Threshold Classifier - MLP $^\dagger$     & & & \checkmark     &63.7 &$-9.3$\\ \hline
     Single Fixed Threshold - 0.5        & \checkmark & &   &44.0 &$-29.0$\\
     Single Best Threshold - 0.1     & \checkmark & & \checkmark     &\underline{62.3} &$-10.7$\\
     Random Sampling - \#Absences=\#Presences            & &\checkmark &     &54.4 &$-18.6$\\
     Random Sampling - \#Absences=5\#Presences           & &\checkmark &     &58.5 &$-14.5$\\
     Random Sampling - \#Absences=10\#Presences          & &\checkmark &     &57.8  &$-15.2$\\
     Random Sampling - 100 Absences                      & &\checkmark &     &55.5 &$-17.5$\\
     Random Sampling - 1000 Absences                     & &\checkmark &     &59.8&$-13.2$\\
     Random Sampling - 10000 Absences                    & &\checkmark &     &45.1&$-27.9$\\
     Target Sampling            & &\checkmark &     &\textbf{62.8} &$-10.2$\\
     Mean Predicted Threshold       & & &     &42.5&$-30.5$\\
     Lowest Presence Threshold (LPT) & & &   &41.6&$-31.4$\\
     Lowest Presence Threshold - Robust (LPT-R)   & & &  &60.2 &$-12.8$\\
    \bottomrule
    \end{tabular}
    }
    \label{tab:more_results_env}
    \vspace{-10pt}
\end{table*} 

\noindent{\bf What impact does the choice of percentile have on \texttt{LPT-R}?}
In \cref{tab:lpt_table} we evaluate how varying the percentiles for the \texttt{LPT-R} method impacts performance across different models and loss functions. 
$\mathcal{L}_{\text{AN-SLDS}}$ seems to perform best with higher percentiles for \texttt{LPT-R} than the other two losses. This implies that there is more `noise' associated with the presences used for identifying the thresholds, \ie more of them need to be discarded when selecting an appropriate threshold. 
These results show that \texttt{LPT-R} can and should be tuned to the specific model it is applied to, ideally using a held-out validation set. 
Note, with the exception of the results in \cref{tab:lpt_table}, this value was set to 5\% by default for all other experiments and is not tuned. 

\begin{table*}[t]
    \centering
    \caption{{\bf Impact of varying the percentile hyperparameter for \texttt{LPT-R}.}
    Here, we report the mean F1 score for different variants of \texttt{LPT-R}, \ie where we use different percentiles when setting the threshold. The scores are computed for the IUCN set as the average of five random initializations of SDMs, both in the setting where the models are only trained on coordinate inputs (`Crds.') and when trained with coordinates together with environmental inputs (`Env. + Crds.'). The scores are presented for different training losses. 
    We can see that the models that also use environmental inputs perform best with a slightly larger percentile over our default of 5\%. 
    }
    \vspace{-7pt}
    \resizebox{0.8\textwidth}{!}{
    \begin{tabular}{l|l|cc}
    \toprule
    Model & Thresholding Method & $\uparrow$ Mean F1 Crds. & $\uparrow$ Mean F1 Env. + Crds.\\
    \midrule 
    $\mathcal{L}_{\text{AN-full}}$ 
         &LPT        &54.3   &41.6\\
         &LPT@2.5    &60.6   &57.0\\
         &LPT@5.0    &\textbf{60.8}   &60.2\\
         &LPT@7.5    &60.0   &\textbf{60.9}\\
         &LPT@10     &58.9   &60.4\\
         &LPT@12.5   &57.6   &59.3\\
         &LPT@15     &56.2   &57.8\\
    \hline
    $\mathcal{L}_{\text{AN-SSDL}}$ 
         &LPT        &49.9   &35.2\\
         &LPT@2.5    &\textbf{53.7}   &50.5\\
         &LPT@5.0    &53.5   &\textbf{53.4}\\
         &LPT@7.5    &52.5   &\textbf{53.4}\\
         &LPT@10     &51.3   &52.5\\
         &LPT@12.5   &50.0   &51.1\\
         &LPT@15     &48.7   &49.5\\
    \hline
    $\mathcal{L}_{\text{AN-SLDS}}$ 
         &LPT        &29.9   &28.9\\
         &LPT@2.5    &36.9   &41.6\\
         &LPT@5.0    &39.4   &47.2\\
         &LPT@7.5    &40.7   &50.7\\
         &LPT@10     &41.5   &52.8\\
         &LPT@12.5   &42.0   &54.2\\
         &LPT@15     &\textbf{42.2}   &\textbf{55.1}\\
    
    \bottomrule
    \end{tabular}
    }
    \label{tab:lpt_table}
    \vspace{-5pt}
\end{table*}

\section{Additional Methods and Baselines}
\label{sec:baselines}
\vspace{-5pt}
Here, we present some additional baselines to help contextualize the performance of the evaluated thresholding techniques. 
These baselines all use the held-out test set directly to find optimal thresholds. 
As a result, this is clearly not a viable approach but serves as a benchmark for the rest of the experiments.

\vspace{3pt}
\noindent{\bf Performance Upper Bound.}
\label{opt ind}
First, we describe how we compute an upper bound on the possible mean F1 scores for the IUCN dataset. 
These values were used to represent the `Upper Bound $\Delta$' in Tab.~\textcolor{red}{2}, where we used the test data to select the optimal threshold for each species. 
These thresholds were obtained by generating predictions for each species for the test locations. Then for each species, the threshold is set to each unique value in the predictions until the one that maximized the F1 score was found. 
This means that each species had its own unique F1 score set. 
The upper bounds obtained are 67.2\%, 60.1\%, and 47.4\% for $\mathcal{L}_{\text{AN-full}}$, $\mathcal{L}_{\text{AN-SSDL}}$, and $\mathcal{L}_{\text{AN-SLDS}}$, respectively. 
These results are the average of five runs with different random initializations of the input SDM. 
The upper bound for the ensemble of five $\mathcal{L}_{\text{AN-full}}$ SDMs is 68.3\%. 
The maximum performance for each of these losses is represented by these scores, meaning that a good thresholding technique would find thresholds that match these as close as possible, and obviously can not be better. 
Similarly, the upper bounds were calculated for the S\&T dataset. 
These scores were 76.1\%, 69.7\%, and 76.2\% for $\mathcal{L}_{\text{AN-full}}$, $\mathcal{L}_{\text{AN-SSDL}}$, and $\mathcal{L}_{\text{AN-SLDS}}$, respectively, and  76.9\% for the ensamble.

\noindent{\bf Subsampling Expert Data.}
We also explore the impact of using a fraction of the high-quality true data to compute the thresholds.
One way of setting the thresholds for species is by using a small amount of true presence-absence data. 
For these experiments on the IUCN dataset, a subsample of the expert evaluation data was used  (1\%, 5\%, and 10\% randomly sampled) to maximize the F1 score and select the threshold. Since the subsample is random, species with a low number of presences might not have any presence-locations at all in the subsample.
These results are presented in \cref{tab:subsample_expert}. 
Although these scores are high and close to the upper bound performance (obtained with 100\% of the data), this method is again not viable for species without expert-derived range maps. 
As can be seen in this experiment, even a small amount of true data enables the identification of almost perfect thresholds for binarizing range maps. 
However, in practice, a sample of even 1\% still requires an infeasible amount of survey locations to be checked. 
For context, the \texttt{LPT-R} approach obtains a mean F1 of 60.8\%, without using pseudo or true absence data, which indicates that it is still quite competitive. 

\begin{table*}[t]
    \centering
    \caption{{\bf Using test data to determine thresholds.} Here we report the mean F1 score for an $\mathcal{L}_{\text{AN-full}}$ SDM when different amounts of evaluation data are used to determine the threshold for each species. A fraction of 1\% indicates that only 1\% of the evaluation presence-absence data is utilized to identify the threshold for each species, and the remaining data is used for evaluation.
    }
    \vspace{-7pt}
    \resizebox{0.4\textwidth}{!}{
    \begin{tabular}{l|c|c}
    \toprule
    Model  & Fraction Used & $\uparrow$ Mean F1\\
    \midrule 
      \multirow{4}{*}{$\mathcal{L}_{\text{AN-full}}$}
            &1\%    &63.7\\
            &5\%    &65.7\\
            &10\%   &66.2\\
            &100\%  &66.4\\
    \bottomrule
    \end{tabular}
    }
    \label{tab:subsample_expert}
    \vspace{-5pt}
\end{table*}

\section{SINR Training Losses}
\label{sec:SINR_losses}
Here we outline the different loss functions used in SINR to train SDMs which are evaluated in Tab.~\textcolor{red}{2} in the main paper.  
SINR introduces the following losses: ``assume negative loss (same species, different location)'' $\mathcal{L}_{\text{AN-SSDL}}$,  ``assume negative loss (same location, different species)'' $\mathcal{L}_{\text{AN-SLDS}}$, and ``full assume negative loss'' $\mathcal{L}_{\text{AN-full}}$. 
The description of the $\mathcal{L}_{\text{AN-full}}$ loss can be found in Sec.~\textcolor{red}{3}. 

The $\mathcal{L}_{\text{AN-SSDL}}$ loss pairs each species observation with a different randomly generated location as a negative (\ie pseudo-absence). 
These randomly generated pseudo-absences are incorporated into the loss function as follows:
\begin{equation}
\mathcal{L}_{\text{AN-SSDL}}(\hat{\mathbf{y}_i}, \mathbf{z}_i) = -\frac{1}{n_{\text{pos}}} \sum_{j=1}^{S}  \mathbbm{1}_{[z_{ij}=1]} \left[ \log(\hat{y}_{ij}) + \log(1 - \hat{y}'_j) \right],
\end{equation}
where a randomly chosen location $\mathbf{r} \sim \text{Uniform}(\mathcal{X})$ is used together with $ n_{\text{pos}} = \sum_{j=1}^{S} \mathbbm{1}_{[z_{ij}=1]}$ to generate $\hat{\mathbf{y}}' = h_{\phi}(f_{\theta}(\mathbf{r}))$. 
In this way random absences are generated across the globe.  

The $\mathcal{L}_{\text{AN-SLDS}}$ loss, on the other hand, associates every species observation with a pseudo-absence at the same location for a different species. 
This generates pseudo-absences that align with the distribution of the presence training data, so is referred to as target background sampling. This loss is computed as:
\begin{equation}
\mathcal{L}_{\text{AN-SLDS}}(\hat{\mathbf{y}_i}, \mathbf{z}_i) = -\frac{1}{n_{\text{pos}}} \sum_{j=1}^{S}  \mathbbm{1}_{[z_{ij}=1]} \left[ \log(\hat{y}_{ij}) + \log(1 - \hat{y}_{ij'}) \right],
\end{equation}
where $j' \sim \text{Uniform(\{} j : z_{ij} \neq 1 \text{\})}$.

%
%
%
\section{Additional Visualizations} 
\begin{figure}[t]
\centering
\includegraphics[width=0.8\linewidth]{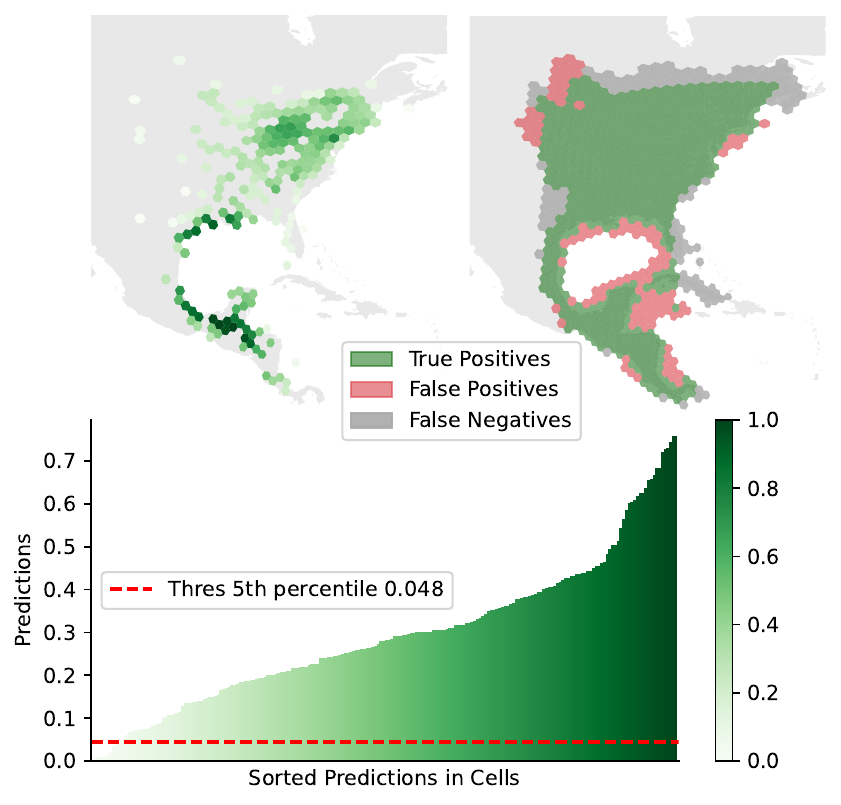}
\vspace{-5pt}
\caption{{\bf Threshold selection for \texttt{LPT-R}.} Here we illustrate the main steps of the \texttt{LPT-R} approach for the Wood Thrush. More specifically, the top left plot shows how the predictions are collected for all of the H3 cells of a specified resolution where the species has been observed. The 5th percentile of these predictions is then calculated and used as a threshold. The map is then binarized so that all cells with a prediction score higher than the threshold are marked as presences. In the top right plot, the output of this process is compared to the expert-derived range map. Green cells indicate true positives, red false positives, and dark gray false negatives.  
}
\label{fig:masking_fig}
\end{figure}

\begin{figure}[h]
\centering
\includegraphics[width=0.9\linewidth]{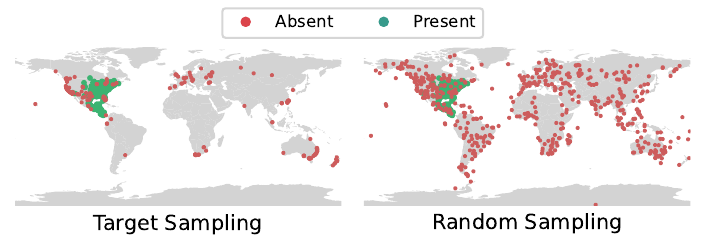}
\vspace{-10pt}
\caption{{\bf Pseudo-absence generation.} Here we visualize how the pseudo-absences (\ie `Absent') are generated for the two common sampling methods, target and random sampling for the Wood Thrush. 
For \texttt{Target Sampling}, the absences are clustered where most of the observations have been reported to iNaturalist, \ie North America, Europe, and parts of Australasia.  
In contrast, the \texttt{Random Sampling} absences are uniformly distributed across the globe.
}
\label{fig:tgt_rnd_comparison}
\vspace{-10pt}
\end{figure}

Finally, we include \cref{fig:masking_fig} and \cref{fig:tgt_rnd_comparison} to visualize how different binarization methods work. 
\cref{fig:masking_fig} illustrates how thresholds are set through \texttt{LPT-R} using the \emph{Wood Thrush} as an example. 
Similarly, this bird species is used to show how pseudo-absences are generated through \texttt{Target Sampling} and \texttt{Random Sampling} in \cref{fig:tgt_rnd_comparison}.

\end{document}